\documentclass[10pt]{revtex4}
\usepackage{hyperref}
\usepackage{amsmath}
\usepackage[psamsfonts]{amssymb }
\usepackage{mathrsfs}

\begin{document}

\title{\Large Weyl Degree of  Freedom in The Nambu-Goto String Through Field Transformation}

\author{Vipul Kumar Pandey\footnote {e-mail address: vipulvaranasi@gmail.com}}
\author{ Bhabani Prasad Mandal\footnote {e-mail address: bhabani.mandal@gmail.com}}

\affiliation {Department of Physics, 
Banaras Hindu University, 
Varanasi-221005, INDIA.}

\begin{abstract}
We show how Weyl degrees of freedom, can be introduced in the Nambu-Goto string in the path integral formulation using reparametrization invariant measure. We first identify Weyl degrees in conformal gauge using BFV formulation. Further we change Nambu-Goto string action to Polyakov action. The generating functional in light-cone gauge is then obtained from the generating functional corresponding to Polyakov action in conformal gauge by using suitably constructed finite field dependent BRST transformation.

\end{abstract}
\maketitle
\section{Introduction}\label{Introduction}

Bosonic strings are formulated using two alternative actions namely Nambu-Goto [NG] action \cite{1} and Polyakov action \cite{2}. In the Polyakov formulation, one uses metric of the string world-sheet, $g_{\alpha\beta}$, to manifest the Weyl and re-parametrization symmetries at the classical level. These symmetries are useful to eliminate the degrees of freedom of  $g_{\alpha\beta}$. Due to the presence of Weyl anomaly, the Weyl freedom may become dynamical at the quantum level \cite{2}. The well known factor, $(D- 26)$, appears naturally in the Polyakov path integral formulation. On the other hand, in the NG string only string coordinates $X^\mu (\mu = 0, ..., D- 1 )$ are used as dynamical variables and hence no Weyl freedom is present from the beginning. Attempts have been made without much success to replace Weyl degrees of freedom by considering quantum longitudinal mode in the covariant formulation of NG string in the sub-critical dimension \cite{3,4}. 
          
 On the other hand non-covariant gauges like light-cone gauge are not easily incorporated in the path integral formulation. Hence theoretical comparison of quantization in various gauges is restricted to certain extent. However this problem is overcomed through BFV formulation \cite{7} by considering the quantization using the most general gauge condition. The main idea of the BFV formulation is to convert the second class constraints to first class constraints, by introducing new canonical variables which are referred as the BF fields. We consider the NG string theory in BFV formulation at the sub-critical dimensions, where the BF fields appear as the conformal degree of freedom \cite{13,14}. The Weyl degrees of freedom is generally introduced based on the analysis of Fujikawa \cite{10} using the re-parametrization invariant  measure for the path integral. It has been observed \cite{10} that for a re-parametrization invariant path-integral measure the string coordinates $X^\mu$ must be scaled by $( - \det g_{\alpha\beta})^\frac{1}{4}$ so as to have the Weyl weight one.
 
 BRST quantization \cite{19} is an important and powerful technique to deal with a system with constraints \cite{20}. It enlarges the phase space of a gauge theory and restores the symmetry of the gauge fixed action in the extended phase space keeping the physical contents of the theory unchanged. We indicate how various BRST invariant effective theories are interlinked by considering the finite-field-dependent version of the BRST(FFBRST) transformation, introduced by Joglekar and Mandal \cite{21} about 24 years ago. FFBRST transformations are the generalization of the usual BRST transformation where the usual infinitesimal, anti-commuting constant transformation parameter is replaced by a field-dependent but global and anti-commuting parameter. Such generalized transformation protects the nilpotency and retains the symmetry of the gauge fixed effective actions. The remarkable property of such transformations are that they relate the generating functional corresponding to different effective actions. The non-trivial Jacobian of the path
integral measure under such a finite transformation is responsible for all the new results. In virtue of this remarkable property, FFBRST transformations have been investigated extensively and have found many applications in various gauge field theoretic systems \cite{21,22,23,24,25,26,27,28,29,30,31,32,33,34}. A similar generalization of the BRST transformation with the same motivation and goal has also been carried out more recently in a slightly different manner \cite{35} where a Jacobian for such transformation is calculated without using any ansatz.

In the present work, we considered Polyakov action in the path integral approach in the conformal gauge when Weyl degrees of freedom are naturally incorporated through reparametrization invariance of path integral measure \cite{10}. Constructing appropriate finite field dependent BRST transformation we connect the generating functional in conformal gauge to that of in light-cone gauge in the Polyakov formulation of NG string. Thus Weyl degrees of freedom is incorporated in the NG string through the FFBRST transformation. 
We will closely follow the procedures followed by Igarashi and Kubo \cite {16}. 

Now we present the plan of the paper. In the next section we will analyse the constraints for NG action and will use BFV formulation \cite{13}. In section III we will transform NG action to Polyakov action. Then BRST invariant actions are written in both  conformal and light-cone gauges. In section IV we will connect generating functionals in conformal and light-cone gauges using appropriately constructed FFBRST transformations. The last section is kept for concluding remarks.

\section{BRST for Nambu-Goto Action}\label{BRST for Nambu-Goto Action}

Nambu-Goto action for the string coordinates $X^{\mu}$, $\mu = 0,1,....D-1$ on a two-dimensional world-sheet parametrized by $x^{\alpha} = (\tau, \sigma)$, $\alpha = 0, 1$ is written as \cite{1} 
\begin{equation}
S_0 = \int d^2x (-\det G_{\alpha\beta})^{\frac{1}{2}}
\label{ngs}
\end{equation}
where
\begin{equation}
G_{\alpha\beta} = \partial_{\alpha}X^{\mu}\partial_{\beta}X_{\mu}
\label{gab}
\end{equation}
The momentum conjugate to $X^\mu$ for this theory is then written as
\begin{equation}
P_\mu = \sqrt {-G} \partial_\alpha X_\mu G^{\alpha 0}
\label{mom}
\end{equation}
where $G = \det G_{\alpha\beta}$. The Hamiltonian corresponding to this system vanishes. This system has two primary constraints which generate two re-parameterizations of string world-sheet and are written as,
\begin{equation}
\phi_{\pm} = \frac {1}{4}(P^2_\mu + (\partial_\sigma X^\mu)^2) \pm\frac {1}{2}\partial_\sigma X^\mu P_\mu 
\label{pmc}
\end{equation}
These constraints are first class at the classical level but appears as  second class at quantum level due to conformal anomaly. To convert them into first class we will introduce new fields $\theta$ and its momentum conjugate $\Pi_\theta$ in the action.
The new effective constraints then take the form 
\begin{equation}
\tilde\phi_{\pm} = \phi_{\pm} + \frac {k}{\sqrt 2}(\partial_\sigma \Pi_\theta \pm (\partial^2_\sigma \theta ) ) + (\Pi_\theta \pm \partial_\sigma \theta)^2 
\label{nec}
\end{equation} 
where $k$ is a constant which is fixed as \cite{9}
\begin{equation}
k = \frac{(25 - D)}{24 \pi}
\label{con}
\end{equation}
We further extend the phase space by introducing following pair of fields 

\begin{equation}
(C^\pm, \bar P_\pm), \quad (P^\pm, \bar C_\pm), \quad (N^\pm, B_\pm )
\label{nvb}
\end{equation}

Now, the action is written in the extended phase space as 
\begin{equation}
S = \int d^2 \sigma [\dot X^\mu P_\mu + \dot \theta \Pi_\theta + \dot C^a \bar P_a + \{\psi, Q \} ]
\label{act}
\end{equation}
where BRST charge $Q$ is given as
\begin{equation}
Q = \int d\sigma\quad[C^\pm (\tilde\phi_{\pm} + \bar P_\pm \partial_\sigma C^\pm ) + B^\pm P_\pm ]
\label{brc}
\end{equation}
One can easily find out that $Q^2$ is nilpotent in nature and gauge-fixing functional takes the form
\begin{equation}
\Psi = \int d\sigma ( i \bar C_a \chi^a + \bar P_a N^a)
\label{gff}
\end{equation}
where $X^a$ does not depend on ghost, anti-ghost, $B$ and $N$ fields.

After eliminating all the non-dynamical variables, BRST transformation for  the dynamical variables is written as \cite{9,13}
\begin{eqnarray}
\delta X^\mu &=& -\frac{1}{2} (C^a \partial_a X^\mu )\nonumber\\
\delta \theta &=& -\frac{1}{2} (C^a \partial_a \theta ) + \frac {k}{2\sqrt 2}(\partial_{+} C^{+} - \partial_{-} C^{-})\nonumber\\
\delta C^\pm &=& -\frac{1}{4} C^\pm \partial_\pm C^\pm \nonumber\\
\delta \bar C_\pm &=& -\frac{1}{4}\partial_\pm X^\mu\partial_\pm X_\mu \pm \bar C_\pm \partial_\pm C^\pm \pm \frac{1}{2}\partial_\pm\bar C_\pm C^\pm - \frac{1}{4}\partial_\pm\theta \partial_\pm\theta \mp \frac {k}{2\sqrt 2}\partial_\pm\partial_\pm \theta
\label{brt}
\end{eqnarray}
which leaves action in Eq. (\ref{act}) invariant.

Now, total Lagrangian density has the form
\begin{equation}
\mathcal{L} = \mathcal {L}_x + \mathcal {L}_{gf} + \mathcal {L}_{gh}
\label{efl}
\end{equation}

where $\mathcal {L}_x$ denote the string part of the Lagrangian density and gauge-fixing and ghost term is defined as 
\begin{equation}
\lambda (\mathcal {L}_{gf} + \mathcal {L}_{gh}) = -i \delta (\bar C_a \chi^a )
\label{gffp}
\end{equation}
where $\lambda$ is infinitesimal Grassmann parameter. Next section we are going to discuss BRST symmetric Polyakov action in conformal as well as in light-cone gauge.

\section{Polyakov Action}\label{Polyakov Action}
 
Following the technique in Ref. \cite{16} we convert NG action to Polyakov action as  
\begin{equation}
\mathcal{L}_x = -\frac{1}{2}\tilde g^{ab}\partial_a X^\mu \partial_b X_\mu - \frac{1}{2}\tilde g^{ab}\partial_a \theta \partial_b \theta
\label{poly}
\end{equation}
Here $\theta$ dependent term in the above Lagrangian density brings extra degrees of freedom in the system. We need a reparametrization invariant measure in the path integral formulation to construct the BRST symmetry of this theory. Using the methods described in \cite{11,15} we construct the BRST transformation as
 
\begin{eqnarray}
\delta X^\mu &=& -\frac{1}{2}(C^a \partial_a X^\mu )\nonumber\\
\delta \theta &=& -\frac{1}{2}(C^a \partial_a \theta ) + \frac {k}{2\sqrt 2}(\partial_0 C^0 - \partial_1 C^1)\nonumber\\
\delta C^a &=& -\frac{1}{4} C^a.
 \partial_a C^a \nonumber\\
\delta \bar C_a &=& -\frac{1}{4}\partial_a X^\mu \partial_a X_\mu \pm \bar C_a \partial_a C^a \pm \frac{1}{2}\partial_a{\bar C_a} C^a - \frac{1}{4}\partial_a\theta \partial_a\theta \mp \frac {k}{2\sqrt 2}\partial_a\partial_a \theta \nonumber\\
\delta \tilde g^{ab} &=& \partial_c C^a \tilde g^{cb} + \partial_c C^b \tilde g^{ac} - \partial_c (C^c \tilde g^{ab}) 
\label{brsp}
\end{eqnarray}

We define the generating functional in path-integral formulation as
\begin{eqnarray}
Z = \int D\phi \exp \big(i\int d^2 x(\mathcal {L}_x + \mathcal {L}_{gf} + \mathcal {L}_{gh})\big)
\label{gf}
\end{eqnarray}
where Lagrangian density is given by Eqs.(\ref{poly}) and (\ref{gffp}) and $D\phi$ is the generic notation for path integral measure. Transformation in Eq.(\ref{brsp}) leaves the effective  action  invariant.
Now we fix the gauge more specifically and discuss BRST invariant effective theories in conformal as well as light cone gauges.

\subsection{Conformal Gauge}\label{Conformal Gauge}

Conformal gauge has been used extensively in the discussion of various problems. It has  been used to study strings, gravity etc in path-integral and covariant operator formalism. This gauge is very useful to remove conformal anomaly, to introduce Weyl symmetry and in renormalizing the theory \cite{9,10,11}.

The conformal gauge condition is expressed as $\tilde g^{ab} = \eta^{ab}$ \cite{11,14}  and is  incorporated into the following gauge-fixing and FP ghost term in a BRST invariant manner,

\begin{eqnarray}
\mathcal{L}_{cf} = \lambda (\mathcal {L}_{gf} + \mathcal {L}_{gh}) = -i \delta^B (\bar C^0\tilde g^{++} + \bar C^1\tilde g^{--})
\label{cfgf}
\end{eqnarray}
Here $\bar C^0$ and $\bar C^1$ are anti-ghost fields.

\subsection{Light-cone Gauge}\label{Light-cone Gauge}

On the other hand, light cone gauge is used to eliminate unphysical degrees of freedom and also in decoupling of ghost fields. Light-cone gauge has  also been used in Kaku-Kikkawa string field theory, in showing the ultraviolet finiteness of $N = 4$ supersymmetric Yang-Mills theory, in dimensional regularization, in gravity, supergravity, string and superstrings theories \cite{18}.

The light-cone gauge condition ,($X^+ = f(\sigma)$, $\tilde g^{++} = 0$) \cite{14,15} is incorporated into the following gauge-fixing and ghost term in a BRST invariant manner, 
\begin{eqnarray}
\mathcal{L}_{lc} = \lambda (\mathcal {L}_{gf} + \mathcal {L}_{gh}) = -i \delta^B (\bar C^0\tilde g^{++} + \bar C^1 (X^+ - f(\sigma ))).
\label{lcgf}
\end{eqnarray}
Here $f(\sigma)$ is an arbitrary function of $\sigma^0$ and $\sigma^1$.

Now we proceed to use FFBRST to address the Weyl degrees of freedom in NG string formulation.

\section{Connection between generating functionals in conformal and light-cone gauges}\label{Connection between generating functionals in conformal and light-cone gauges}

Before going for the connection between the two effective theories we briefly discuss the ideas of FFBRST developed in Ref.\cite{21}. The BRST transformations  are generated from BRST charge using relation $\delta_b \phi = -[Q,\phi]\delta\Lambda$ where $\delta\Lambda$ is infinitesimal anti-commuting global parameter. Following their technique  the anti-commuting BRST parameter $\delta\Lambda$ is generalized to be finite-field dependent instead of infinitesimal but space time independent parameter $\Theta[\phi]$. Since the parameter is finite in nature unlike the usual case  the path integral measure is not invariant under such finite transformation.  The Jacobian for these transformations for certain $\Theta[\phi]$ can be calculated by following way.  
\begin{eqnarray}
D\phi &=& J(k)D\phi'(k)\nonumber\\
       &=& J(k+dk)D\phi'(k+dk)
\label{chj}
\end{eqnarray}
where  a  numerical parameter $k$ ($0\leq k \leq 1$), has been introduced  to execute the finite transformation in a mathematically convenient way. All the fields are taken to be $k$ dependent in such a fashion that $\phi(x,0) = \phi(x)$ and $\phi(x,k = 1) = \phi'(x)$. 
$S_{eff}$ is invariant under FFBRST which is constructed by considering successive infinitesimal BRST transformations $(\phi(k)\rightarrow\phi(k+dk))$. The nontrivial Jacobian $J(k)$ is then written as local functional of fields and will be replaced as $e^{iS_1[\phi(k),k]}$ if the condition 

\begin{eqnarray}
\int D\phi(k)\big[\frac{1}{J(k)}\frac{d J(k)}{d k}-i\frac{dS_1}{dk}\big] e^{i (S_1 + S_{eff})} = 0 
\label{ffbc}
\end{eqnarray}
holds \cite{20}. Where $\frac{dS_1}{dk}$ is a total derivative of $S_1$ with respect to $k$ in which dependence on $\phi(k)$ is also differentiated. The change in Jacobian is calculated as 
\begin{eqnarray}
\frac{J(k)}{J(k+dk)} &=& \Sigma_{\phi}{\pm}\frac{\delta \phi(x,k+dk)}{\delta \phi(x,k)}\nonumber\\
                     &=& 1 - \frac{1}{J(k)}\frac{d J(k)}{d k} d k 
 \label{echj}
\end{eqnarray}
${\pm}$ is for bosonic and fermionic fields respectively.

In this section, we construct the FFBRST transformation with an appropriate finite parameter to obtain the generating functional corresponding to $\mathcal{L}_{cf}$ from that of  corresponding to $\mathcal{L}_{lc}$. We calculate the Jacobian corresponding to such a FFBRST transformation following the method outlined in Ref. \cite{21} and show that it is a local functional of fields and accounts for the differences of the two FP effective actions.

The generating functional corresponding to the FP effective action $S_{cf}$ is written as
\begin{eqnarray}
Z_{cf} = \int D\phi \exp (iS_{cf}[\phi])
\label{gfcfg}
\end{eqnarray}
where $S_{cf}$ is given by
\begin{eqnarray}
S_{cf} = \int d^2 x(\mathcal {L}_x + \mathcal {L}_{cf})
\label{scf}
\end{eqnarray}
Now, to obtain the generating functional corresponding $S_{lc}$, we  apply the FFBRST transformation with a finite parameter $\Theta[\phi]$ which is obtained from the infinitesimal but field dependent parameter, $\Theta'[\phi(k)]$ through $ \int_0^\kappa \Theta^\prime[\phi(\kappa)] d\kappa$, we construct $ \Theta^\prime[\phi(\kappa)] $ as,
\begin{eqnarray}
\Theta'[\phi] = i \int d^2 x [\gamma\bar C^1 \{ (X^+ - f(\sigma )) - \tilde g^{--}\} ]
\label{thd}
\end{eqnarray}
Here $\gamma$ is arbitrary constant parameter and all the fields depend on the parameter $k$.
The infinitesimal change in the Jacobian corresponding to this FFBRST transformation is calculated using Eq.(\ref{echj}) 
\begin{eqnarray}
 \frac{1}{J(k)}\frac{d J(k)}{d k} = -i \int d^2 x \gamma[ \delta^B (\bar C^1)\{ (X^+ - f(\sigma )) - \tilde g^{--}\} - (C^a \partial_a X^+ )\bar C^1 - \delta\tilde g^{--}\bar C^1] 
 \label{chij}
\end{eqnarray}
To express the Jacobian contribution in terms of a local functional of fields, we make an ansatz for $S_1$ by considering all possible terms that could arise from such a transformation as
\begin{eqnarray}
S_1[\phi(k), k]  &=&  \int d^2 x [- \xi_1 \delta^B(\bar C^1) (X^+ - f(\sigma )) - \xi_2 \delta^B(\bar C^1)\tilde g^{--} +\xi_3(C^a \partial_a X^+ )\bar C^1 + \xi_4\delta\tilde g^{--}\bar C^1 \nonumber\\ &-& \xi_5 \delta^B(\bar C^0)\tilde g^{++} + \xi_6 \bar C^0 \delta\tilde g^{++}] 
\label{son}
\end{eqnarray}
where all the fields are considered to be $k$ dependent and we have introduced arbitrary $k$ dependent parameters $\xi_n=\xi_n(k) (n =1, 2, .., 6)$ with initial condition $\xi_n(k = 0) = 0$. It is straight forward to calculate
\begin{eqnarray}
\frac{dS_1}{dk} &=& \int d^2 x [ -\xi'_1 \delta^B(\bar C^1) (X^+ - f(\sigma )) - \xi'_2 \delta^B(\bar C^1)\tilde g^{--} +\xi'_3 (C^a \partial_a X^+ )\bar C^1 + \xi'_4\delta\tilde g^{--}\bar C^1 - \xi'_5 \delta^B(\bar C^0)\tilde g^{++}\nonumber\\ &+& \xi'_6 \bar C^0 \delta\tilde g^{++} + \Theta'\{\xi_1 C^a \partial_a X^+ \delta^B(\bar C^1) - \xi_2 \delta\tilde g^{--}\delta^B(\bar C^1) - \xi_3 (\delta^B(\bar C^1))C^a \partial_a X^+ + \xi_3 (C^b \partial_b C^a)\partial_a X^+ \bar C^1 \nonumber\\ &+& \xi_3 C^a \partial_a (-C^b \partial_b X^+)\bar C^1 + \xi_4 \delta^B(\bar C^1)\delta\tilde g^{--} - \xi_5 \delta\tilde g^{++} \delta^B(\bar C^0) + \xi_6 \delta^B(\bar C^0) \delta\tilde g^{++}\} ] 
\label{dsdk}
\end{eqnarray} 

where $\xi'_n = \frac {d\xi_n}{dk}$.
Now we will use the condition of Eq.(\ref{ffbc}).
\begin{eqnarray}
 &\int  D\phi &  \exp[{i (S_{cf}[\phi(k) ] + S_1[\phi(k), k] )}]\int d^2 x [(\gamma - \xi'_1 ) \delta^B (\bar C^1) (X^+ - f(\sigma )) - (\gamma + \xi'_2 ) \delta^B (\bar C^1)\tilde g^{--} \nonumber\\ &+&(-\gamma + \xi'_3 ) (C^a \partial_a X^+ )\bar C^1 + (-\gamma + \xi'_4 )\delta\tilde g^{--}\bar C^1 - \xi'_5 \delta^B (\bar C^0)\tilde g^{++} + \xi'_6 \bar C^0 \delta\tilde g^{++} \nonumber\\ &+& \Theta'\{(\xi_1 - \xi_3 ) C^a \partial_a X^- \delta^B (\bar C^1) - (\xi_2 +  \xi_4 )\delta\tilde g^{--}\delta^B (\bar C^1) + (-\xi_5 + \xi_6)\delta\tilde g^{++} \delta^B (\bar C^0) \} ] = 0
 \label{ffbrc}
\end{eqnarray} 

The terms proportional to $\Theta'$, which are nonlocal due to $\Theta'$, vanish independently  if

\begin{eqnarray}
\xi_1 - \xi_3 &=&0\nonumber\\
\xi_2 + \xi_4 &=&0\nonumber\\
-\xi_5 + \xi_6 &=&0
\label{reiji}
\end{eqnarray}
To make the remaining local terms in Eq.(\ref{ffbrc}) vanish, we need the following conditions:

\begin{eqnarray}
\gamma - \xi'_1 &=&0\nonumber\\
\gamma + \xi'_2 &=&0\nonumber\\
\gamma - \xi'_3 &=&0\nonumber\\
\gamma - \xi'_4 &=&0\nonumber\\
\xi'_5 &=&0\nonumber\\
\xi'_6 &=&0
\label{rbjg}
\end{eqnarray}
The differential equations for $\xi_n(k)$ can be solved with the initial conditions $\xi_n(0) = 0$ to obtain the solutions
\begin{eqnarray}
\xi_1 = \gamma k, \quad  \xi_2 = -\gamma k,\quad \xi_3 = \gamma k,\quad \xi_4 = \gamma k,\quad \xi_5 = \xi_6 = 0 
\label{voj}
\end{eqnarray}

Putting values of these parameters in expression of $S_1$, and choosing arbitrary parameter $\gamma = -1$ we obtain,
\begin{eqnarray}
S_1[\phi(1), 1] = \int d^2 x [ \delta^B(\bar C^1) (X^+ - f(\sigma )) - \delta^B{\bar C^1}\tilde g^{--} - (C^a \partial_a X^+ )\bar C^1 - \delta\tilde g^{--}\bar C^1 ] 
\label{yeso}
\end{eqnarray}
Thus the FFBRST transformation with the finite parameter $\Theta$ that is defined by Eq.(\ref{thd}) changes the generating functional $Z_{cf}$ as
\begin{eqnarray}
Z_{cf} &=& \int D\phi \exp (iS_{cf}[\phi] )\nonumber\\
&=&\int D\phi'\exp[{i (S_{cf}[\phi'] + S_1[\phi', 1] )}] \nonumber\\
&=&\int D\phi\exp[{i (S_{cf}[\phi] + S_1[\phi , 1] )}] \nonumber\\
&=& \int D\phi \exp (iS_{lc}[\phi] ) \equiv Z_{lc}
\label{gflc}
\end{eqnarray}
Here $S_{lc}$ is defined as
\begin{eqnarray}
S_{lc} = \int d^2 x(\mathcal {L}_x + \mathcal {L}_{lc})
\label{slc}
\end{eqnarray}
In this way FFBRST transformation with the finite field dependent parameter in Eq. \ref{thd} connects generating functional for the Polyakov action in conformal gauge to that of in the light-cone gauge. 
\section{Conclusion}\label{Conclusion}
In this present work we have demonstrated how Weyl degree of freedom are incorporated in the formulation of NG string through certain field transformation. Weyl degree of freedom are first identified in conformal gauge using BFV formulation. Then we have established the connection between conformal gauge to light-cone in Polyakov type action for NG string using the technique of FFBRST transformation, which connects various theories through the non-trivial Jacobian of path integral measure. The non-local BRST transformation by Igarashi etal in \cite{16} is nothing but a particular type of FFBRST transformation. The parameter $\lambda$ in the non-local transformation in \cite {16} is identified with FFBRST parameter $\Theta'$.

One of us (VKP) acknowledges the University Grant Commission (UGC), India, for its financial assistance under the CSIR-UGC JRF/SRF scheme.


\begin{thebibliography}{99}
\bibitem{1}Y. Nambu, lectures at the Copenhagen Symp. (1970); T. Goto, Prog. Theor. Phys., 46 (1971) 1560.
\bibitem{2}A. M. Polyakov, Phys. Lett. B, 103 ( 1981 ) 207.
\bibitem{3}P. Goddard, C. Rebbi and C. B. Thorn, Nuov. Cimento, 12(1972) 425; A. Patrascioiu, Nucl. Phys. B, 81 (1974) 525; W. A. Bardeen, 1. Bars, A. J. Hanson and R. D. Peccei, Phys. Rev. D, 13 (1976) 2364.
\bibitem{4}E. S. Fradkin and A. A. Tseytin, Ann. Phys., 143 (1982) 413.
\bibitem{5}L. D. Faddeev and V. N. Popov, Phys. Lett. B, 25 (1967) 29.
\bibitem{6}L. D. Faddeev, Theor, Math. Phys., 1 (1976) 1.
\bibitem{7}E. S. Fradkin and G. Vilkovisky, Phys. Lett. B, 55 (1975) 224; I. A. Batalin and G. Vilkovisky, Phys. Lett. B, 69 (1977) 309; I. A. Batalin and E. S. fradkin, Phys. Lett. B, 122 (1983) 157.
\bibitem{8}P. A. M. Dirac, Proc. R. Soc. London A, 246 (1958) 326; Lectures on quantum mechanics (Yeshiba Univ. Press, New York, 1964).
\bibitem{9}S. Hwang, Phys. Rev. D, 28 (1983) 2614; K. Fujikawa, T. Inagaki and H. Suzuki, Phys. Lett. B, 213 (1988) 279.
\bibitem{10}K. Fujikawa, Phys. Rev. D, 25 (1982) 2584; K. Fujikawa, Nucl. Phys. B, 291 (1987) 583.
\bibitem{11}M. Kato and K. Ogawa, Nuclear Physics B, 212 (1983) 443.
\bibitem{12}K. Fujikawa, U. Lindstrom, N.K. Nielsen, M. Rocek and P. van Nieuwenhuizen, Phys. Rev. D, 37 (1988) 391. 
\bibitem{13}T. Fujiwara, Y. Igarashi and J. Kubo, Nuclear Physics B, 341 (1990) 695.
\bibitem{14}T. Fujiwara, Y. Igarashi, J. Kubo and T. Tabei, Phys. Rev D, 48 (1993) 1736.
\bibitem{15}R. Tzani, Phys. Rev. D, 43 (1991) 1254.
\bibitem{16}Y. Igarashi and J. Kubo, Phys. Lett. B, 217 (1989) 55.
\bibitem{17}E. Witten, Nucl. Phys. B, 268 (1986) 253.
\bibitem{18}M. Kaku and K. Kikkawa, Phys. Rev. D, 10 (1974) 1110, G. Leibbrandt Reviews of Modern Physics, 59 (1987) No. 4.
\bibitem{19}C. Becchi, A. Rouet and R. Stora, Phys. Lett. B, 52 (1974) 344; Ann. Phys. (N. Y.) 98 (1976) 287; I. V. Tyutin, Lebedev Report N FIAN 39, 1975 (unpublished).
\bibitem{20}M. Henneaux, �Hamiltonian form of the path integral for theories with a gauge freedom,� Physics Reports, vol. 126, no. 1 (1985) pp. 1�66; M. Henneaux and C. Teitelboim, Quantization of Gauge Systems, Princeton University Press, 1992; K. Sundermeyer, Constrained Dynamics, vol. 169 of Lecture Notes in Physics, Springer, Berlin, Germany, 1982. 
\bibitem{21}S. D. Joglekar and B. P. Mandal, Phys. Rev. D, 51 (1995) 1919.
\bibitem{22}S. D. Joglekar and A. Misra, J. Math. Phys 41, 1755,(2000). Int. J. Mod. Phys. A, 15 (2000);Mod. Phys. Lett. A, 14(1999) 2083; Mod. Phys. Lett. A 15,(2000) 541; S. D. Joglekar, Mod. Phys. Lett A, 15(2000) 245.
\bibitem{23}S. D. Joglekar and B. P. Mandal, Int. J. Mod. Phys. A, 17 (2002) 1279. 
\bibitem{24}S. D. Joglekar, Int. J. Mod. Phys. A, 16 (2000) 5043.
\bibitem{25}R. Banerjee and B. P. Mandal, Phys. Lett. B, 27(2000) 488.
\bibitem{26}R. S. Bandhu and S. D. Joglekar, J. Phys. A, 31(1998) 4217.
\bibitem{27} S K Rai and B P Mandal, IJTP, 52,(2013) 3512.
\bibitem{28} S K Rai and B P Mandal, EPJC, 63(2009) 323.
\bibitem{29}S. Upadhyay and B. P. Mandal, EPL, 93 (2011) 31001.
\bibitem{30}S. Deguchi, V. K. Pandey and B. P Mandal  Phys. Lett. B, 756 (2016) 394.
\bibitem{31}S. Upadhyay and B. P. Mandal, Phys. Lett. B, 744(2015) 231.
\bibitem{32}S. Upadhyay, A. Reshetnyak and B. P. Mandal , Eur. Phys. J. C, 76 (2016) 391.
\bibitem{33}V. K. Pandey and B. P. Mandal, Euro. Phys Lett., 119 (2017) 31003
\bibitem{34}V. K. Pandey and B. P Mandal, Advances in High Energy Physics, 10(2017) 1155.
\bibitem{35}P. M. Lavrov and O. Lechtenfeld, Phys. Lett. B, 725(2013) 382.; P. Y. Moshin and A. A. Reshetnyak, Phys. Lett. B, 739 (2014) 110;P. Y. Moshin and A. A. Reshetnyak, Nucl. Phys. B, 888 (2014) 92.

\end{thebibliography}
\end{document}